\documentclass[twocolumn,tighten,trackchanges]{aastex63}
\usepackage[caption=false]{subfig}

\turnoffedit
\usepackage{afterpage}

\submitjournal{ApJ}

\shorttitle{AGN feedback in NGC~3079}
\shortauthors{Fernandez et al.}

\begin{document}

\title{FRAMEx~IV: Mechanical Feedback from the Active Galactic Nucleus in NGC~3079}

\correspondingauthor{Luis C. Fernandez}
\email{lfernan@gmu.edu}

\author[0000-0002-0819-3033]{Luis C. Fernandez}
\affiliation{Department of Physics and Astronomy, George Mason University, MSN 3F3, 4400 University Dr, Fairfax, VA 22030-4444, USA}

\author[0000-0002-4902-8077]{Nathan J. Secrest}
\affiliation{U.S. Naval Observatory, 3450 Massachusetts Ave NW, Washington, DC 20392-5420, USA}

\author[0000-0002-4146-1618]{Megan C. Johnson}
\affiliation{U.S. Naval Observatory, 3450 Massachusetts Ave NW, Washington, DC 20392-5420, USA}

\author[0000-0002-3365-8875]{Travis C. Fischer}
\affiliation{AURA for ESA, Space Telescope Science Institute, 3700 San Martin Drive, Baltimore, MD 21218, USA}

\begin{abstract}
Using the Very Long Baseline Array, we observed the active galactic nucleus (AGN) in NGC~3079 over a span of six months to test for variability in the two main \edit1{parsec-scale } radio components, \textit{A} and \textit{B}, which lie on either side of the AGN. We found evidence for positional differences in the positions of \textit{A} and \textit{B} over the six months consistent with the apparent motion of these components extrapolated from older archival data, finding that their projected rate of separation, \edit1{$(0.040\pm0.003)$~c}, has remained constant since $\sim2004$ when a slowdown concurrent with a dramatic brightening of source \textit{A} occurred. This behavior is consistent with an interaction of source \textit{A} with the interstellar medium (ISM), as has previously been suggested in the literature. We calculated the amount of mechanical feedback on the ISM for both the scenario in which \textit{A} is an expulsion of material from the central engine and the scenario in which \textit{A} is a shock front produced by a relativistic jet, the latter of which is favored by several lines of evidence we discuss. We find that the cumulative mechanical feedback on the ISM is between $2 \times 10^{44}$~erg to $1 \times 10^{48}$~erg for the expulsion scenario or between $3\times 10^{50}$~erg to $1 \times 10^{52}$~erg for the jet scenario. Integrated over the volume-complete FRAMEx sample, 
our results imply that jet-mode mechanical feedback plays a negligible role in the energetics of AGNs in the local universe.


\end{abstract}


\section{Introduction} \label{section:introduction}

\begin{deluxetable*}{llccccccc}
\tablecaption{\\
VLBA Observations of NGC~3079 \label{tab:radio_obs}}

\tablehead{   & \colhead{T$_{\rm int}$} & \colhead{F$_{\rm center}$} &               \colhead{Bandwidth} & \colhead{F$_{\rm range}$} & \colhead{Restoring Beam} &                \colhead{Beam angle} & \colhead{RMS} & \colhead{RMS$_\mathrm{theoretical}$}\\  [-0.2cm]
            \colhead{Date} &\colhead{(s)} &\colhead{(GHz)} & \colhead{(MHz)} & \colhead{(GHz)} & \colhead{($\alpha \times \delta$; mas)}	& \colhead{(deg)} & \colhead{($\mu$Jy bm$^{-1}$)} & \colhead{($\mu$Jy bm$^{-1}$)} }

\startdata
Not Tapered\\
27 Jan 2020 & 3488 & 5.800318 & 384 & 5.612$-$5.996 & 4.46$\times$3.51 & -45.6 & 94 & 19\\
19 May 2020 & 3500 & 5.803460 & 384 & 5.612$-$5.996 & 4.01$\times$2.57 & -46.3 & 75 & 19\\
\hline
Tapered\\
27 Jan 2020 & 3488 & 5.799807 & 384 & 5.612$-$5.996 & 6.69$\times$5.48 & -25.3 & 161 & 19\\
19 May 2020 & 3500 & 5.803971 & 384 & 5.612$-$5.996 & 6.57$\times$5.19 & -33.1 & 199 & 19\\
\enddata
\tablecomments{The calibrator J0956+5753 used for phase referencing. For tapered images to equal $uv$ length of 48 M$\lambda$}

\end{deluxetable*}

\begin{deluxetable*}{lccccc}
\tablecaption{\\
6 GHz VLBA Measurements of NGC~3079 \label{tab:radio_measurements}}

\tablehead{\colhead{Observation Date}	& \colhead{$I_\mathrm{\nu ~peak}$}  &                                \colhead{$F_\mathrm{peak}$} & \colhead{log$_{10}(L_\mathrm{peak}\  / \mathrm{\ erg} \mathrm{\ s}{^{-1}})$} &                            \colhead{$S_\mathrm{int}$}  & \colhead{Noise RMS}\\  [-0.2cm] &
           \colhead{(mJy~beam{$^{-1}$})}	& \colhead{({${\times}10^{-16}$} erg~s{$^{-1}$} cm{$^{-2}$})} 	&  &   \colhead{(mJy)} & \colhead{(mJy)}
            }
\startdata
Component \text{A}\\\hline
\\
Untapered \\
27 Jan 2020             & 38$\pm$2  & 22$\pm$1 & 37.82     & 64$\pm$3  & 0.094\\
19 May 2020             & 36$\pm$2  & 21$\pm$1 & 37.80     & 58$\pm$3  & 0.075\\
\\
Tapered\\
27 Jan 2020             & 49$\pm$2  & 28$\pm$1 & 37.93     & 62$\pm$3  & 0.161\\
19 May 2020             & 46$\pm$2  & 26$\pm$1 & 37.90     & 56$\pm$3  & 0.199\\
\\
Component \text{B}\\\hline
\\
Untapered \\
27 Jan 2020             & 6.7$\pm$0.3   & 3.9$\pm$0.2 & 37.07      & ~9.5$\pm$0.5  & 0.094\\
19 May 2020             & 8.6$\pm$0.4   & 5.0$\pm$0.2 & 37.18      & 11.0$\pm$0.6  & 0.075\\
\\
Tapered\\
27 Jan 2020             & 7.0$\pm$0.4  & 4.1$\pm$0.2 & 37.09     & 7.7$\pm$0.5  & 0.161\\
19 May 2020             & 8.7$\pm$0.4  & 5.0$\pm$0.3 & 37.18     & 9.6$\pm$0.6  & 0.199\\
\\
Component \text{C}\\\hline
27 Jan 2020             & 2.0$\pm$0.1   & 1.17$\pm$0.08 & 36.55      & 5.1$\pm$0.4  & 0.094\\
19 May 2020             & 1.9$\pm$0.1   & 1.12$\pm$0.07 & 36.53      & 3.8$\pm$0.3  & 0.075\\
\enddata
\tablecomments{Measured peak and integrated flux values with 1$\sigma$ uncertainties calculated using the task \textsc{jmfit} which uses an elliptical Gaussian fitting algorithm. There is an additional uncertainty of 5\% of the measured flux added in quadrature to the $1\sigma$ to account for the absolute flux uncertainty of the VLBA. Measurements for tapered are from a $uv$ coverage of equal distance of 48~M$\lambda$.}
\end{deluxetable*}

The Fundamental Reference AGN  Monitoring Experiment \citep[FRAMEx;][]{2020jsrs.conf..165D}, led by the U.S. Naval Observatory, is an ongoing campaign to better understand the physical mechanisms that can affect the apparent positions and morphologies of active galactic nuclei (AGNs) as a function of wavelength. In FRAMEx~I \citep[][]{2021ApJ...906...88F}, we observed a volume-complete ($D<$~40~Mpc) sample of 25 nearby AGNs with a snapshot campaign using simultaneous observations with the Very Long Baseline Array (VLBA) and Swift X-ray Telescope (XRT). We found that the ``fundamental plane'' of black hole activity \citep[e.g.,][]{2003MNRAS.345.1057M}, which purports to unify the X-ray and radio luminosities of AGNs and X-ray binaries through the black hole mass, breaks down  at high physical resolution. Moreover, despite all FRAMEx AGNs having hard X-ray ($14-195$~keV) luminosities larger than $10^{42}$~erg~s$^{-1}$ by construction, only nine out of the 25 AGNs have detectable 6\,GHz radio emission down to a depth of $20$\,$\mu$Jy at $\sim3$~mas (sub-parsec) spatial scales. To explore the role of variability in the sub-parsec regime, we followed FRAMEx~I with a 6-month VLBA and Swift~XRT campaign that observed the 9 detected sources in a 28~day cadence. FRAMEx~II \citep[][]{2022ApJ...927...18F} presented the results for NGC~2992 and found anti-correlated radio and X-ray variability that is consistent with an outburst from the accretion disk simultaneously increasing the free-free absorption depth and the number of electrons available for inverse-Compton scattering of UV photons. The results of the six-month campaign for the remaining FRAMEx AGNs are forthcoming. FRAMEx~III \citep[][]{2022ApJ...936...76S} explored the radio non-detections from FRAMEx~I using the VLBA with longer integration times, expanding the original sample with an additional 9 objects that have redshift-independent distances consistent with our volume definition to improve completeness. We found that, despite an observation depth of $8$\,$\mu$Jy, the majority of the sample remained undetected at mas spatial scales, although five new detections were recorded bringing the total detection fraction to 14/34 (41\%). The X-ray-based radio-loudness parameter $R_X\equiv L_R / L_X$ of these extremely radio-faint AGNs showed an anti-correlation with Eddington ratio, similar to the behavior found in X-ray binaries.

Continuing with the goals of FRAMEx, a major field of ongoing research is examining how AGN ``feedback'' affects host galaxies from their immediate surroundings at parsec scales, upwards to kpc scales, and in rare instances Mpc scales due to jets, outflows, or a combination of both. \edit1{Quantifying on which scales} feedback occurs is critical to understanding the relationship between AGN activity and star formation, heating of the inter-cluster medium, and the co-evolution of supermassive black holes (SMBHs) and their host galaxies more generally. \citet[][]{2008ASPC..386..240L} discusses some of the basic forms of jet interactions with the surrounding medium at parsec scales where the jet is susceptible to external interactions, falling into three classifications: bow shock-hotspot interactions, cloud collisions, and entrainment. With a sufficiently long temporal baseline of very long baseline interferometry (VLBI) observations, the generation and evolution of these structures can be seen directly, allowing for quantification of feedback mechanisms on parsec scales.

Since the early 1980s, the AGN in NGC~3079 has been observed multiple times using VLBI \citep[e.g.][]{1982A&A...114..400H,1988ApJ...335..658I,1990ApJ...356..149H}, including several multi-frequency campaigns to probe the core components of the AGN \citep[e.g.][]{1998ApJ...495..740T,2000PASJ...52..421S,2005ApJ...618..618K}. \cite{1998ApJ...495..740T} observed NGC~3079 with the VLBA at 5 and 8~GHz in 1992 and 22~GHz in 1995 where they observed two compact sources (components \textit{A} and \textit{B}). They also detected a third component (\textit{C}) that lies along the same axis as \textit{A} and \textit{B} (their Fig. 5) at 22~GHz, unfortunately they were unable to detect component \textit{C} at 5 or 8~GHz. At 22~GHz, the emission from \textit{B} is the dominating feature with an integrated flux of $\sim$16~mJy. \cite{1998ApJ...495..740T} stated that none of the sources detected appeared to mark the nuclear engine, which was inferred from the presence of water maser emission, and may only represent features of a nuclear jet. \cite{2000PASJ...52..421S} observed NGC~3079 with the VLBA at 1.4, 8.4, 15, and 22~GHz in 1996. Due to the restoring beam of the VLBA at 1.4~GHz, components \textit{A} and \textit{C} appear as a single component, while at 8.4~GHz the components \textit{A}, \textit{B}, and \textit{C} are resolved. Only the \textit{B} component was detected at 15 and 22~GHz (their Fig. 1).  
In addition to these results, these studies have found the brighter off-nuclear components \textit{A} and \textit{B}, to be increasing in separation. \cite{2007MNRAS.377..731M} observed the separating nuclear components from 1999 to 2005 in C-band at 5 GHz. Using their observations with archival data, they found the rate of separation of components \textit{A} and \textit{B} to be declining. They attributed the slow down due to collisions with the interstellar medium. \cite{2007ApJ...658..203H} observed NGC~3079 in 2006 and suggested the radio knots could be a compact supermassive binary black hole system, but \cite{2016MNRAS.459..820T} determined this was not the case due to NGC~3079 not having two distinct compact radio cores that have a flat or inverted spectrum. Component \textit{A} has changed from steep to flat spectrum while component \textit{B} has remained inverted. 
 
Since 2006, there have been only a few VLBA observations of NGC~3079, most of which have either instrumental effects, limited $uv$ coverage, or low integration times that preclude accurate measurements of the positions and flux densities of components \textit{A} and \textit{B}. {There were three C-band observations and all three datasets contained issues. The data from an observation in 2016 had a low integration time and limited $uv$ coverage, causing a larger restoring beam where multiple nuclear components appear as a single component. A 2019 observation, from the FRAMEx~I snapshot, also suffered from similar issues where the data suffered from instrumental effects and only a single component was observed. We also examined the EVN archive and found two observations, one in February 2019 and one in October 2019. The observation in October requested time to study the absorption line of OH at 6~GHz and not continuum. The results from these observations have yet to be published. This means there has not been any accurate measurements of the ongoing separation of the nuclear components \textit{A} and \textit{B} for $\sim$12 years.} Considering that the AGN in NGC~3079 has been observed previously multiple times over $\sim40$ years, this provides an opportunity to study the interaction of the AGN with the surrounding medium. In this work, we use the VLBA observations from our six-month campaign to dramatically increase the baseline of VLBI observations of the AGN in NGC~3079. From morphological and radio-loudness considerations, we argue that the radio structures in the inner few parsec of NGC~3079 are jet-powered, although we consider an alternative, outflow scenario, and we estimate the amount of mechanical (kinetic) energy deposited on the interstellar medium (ISM) by radio-mode feedback in both cases. 



\section{Methodology}\label{sec: Methodology}
As in \cite{2021ApJ...906...88F}, we used a redshift of $z=0.0037$ and a distance $D = 15.9$~Mpc for NGC~3079, which translates to an angular scale of 0.077~pc~mas$^{-1}$.  

\subsection{\textit{Very Long Baseline Array} Observations} \label{subsection: VLBA}

Through the U.S.\ Naval Observatory's 50\% timeshare, we received observation time on the VLBA at 5 cm (6 GHz) every 28 days (PI:~T.\ Fischer). This began on December 31, 2019 and provided a total of 6 observations. We \edit1{followed} the same phase referencing method described in \citet{2021ApJ...906...88F} and \citet{2022ApJ...927...18F}. For our \edit1{time-series observations}, we requested an integration time of 1~hour with all 10 antennas for NGC~3079. Unfortunately, on 12/31/2019 and 04/21/2020 not all 10 antennas were used. On 12/31/2019 antennas HN and OV did not participate and on 04/21/2020 antenna MK did not participate. From previous measurements \citep[][]{2007MNRAS.377..731M}, the separation rate of the nuclear components \textit{A} and \textit{B} indicate that the observable separation will be difficult to distinguish at our cadence. Therefore, \edit1{we present only the first and last} observations from the time series that utilized all 10 antennas to obtain the highest flux sensitivity and consistent angular resolution between epochs. Table~\ref{tab:radio_obs} summarizes these observations.

\subsubsection{Calibration \& Imaging}

We followed the same steps described in \citet{2022ApJ...927...18F} to calibrate and image the VLBA data of NGC~3079 using NRAO's software package, Astronomical Image Processing System (AIPS) \citep{2003ASSL..285..109G}, release \texttt{31DEC19}. These steps corrected for ionospheric delays, Earth orientation parameters, sample threshold errors, instrument delays, bandpass, amplitude, and parallactic angle. We flagged radio frequency interference in both time and by frequency using the tasks \textsc{editr} and \textsc{wiper} respectively. Next, we calibrated the phase and absolute astrometry of our data to the accuracy of the phase calibrator's position using a two-point interpolation function. The phase calibrator used for NGC~3079 is ICRF~J095622.6+575355, which has R.A. and decl.\ of $\alpha= 149\fdg09431037(8)$, $\delta = 57\fdg89886214(3)$ in ICRF3 \citep[][]{2020A&A...644A.159C}, where the parentheses denote the uncertainties (on the order of $100~\mu \mathrm{as}$).

To image the calibrated data, we followed the same steps outlined in \citet{2022ApJ...927...18F} using the AIPS task \textsc{imagr}. Since the \edit1{data have} high S/N, we were able to apply self-calibration to the images by first self-calibrating on phase only, and then on a combination of phase and amplitude using the AIPS task \textsc{calib}. This was an iterative process where we used the new image as a model for the subsequent calibration until the thermal RMS noise could not be improved any further without introducing artifacts and falling below theoretical RMS. The final self-calibrated images were then used to analyze the flux densities of components \textit{A} and \textit{B} (we also include the results for component \textit{C}).  

\begin{figure*}
    \centering\noindent
    \gridline{\fig{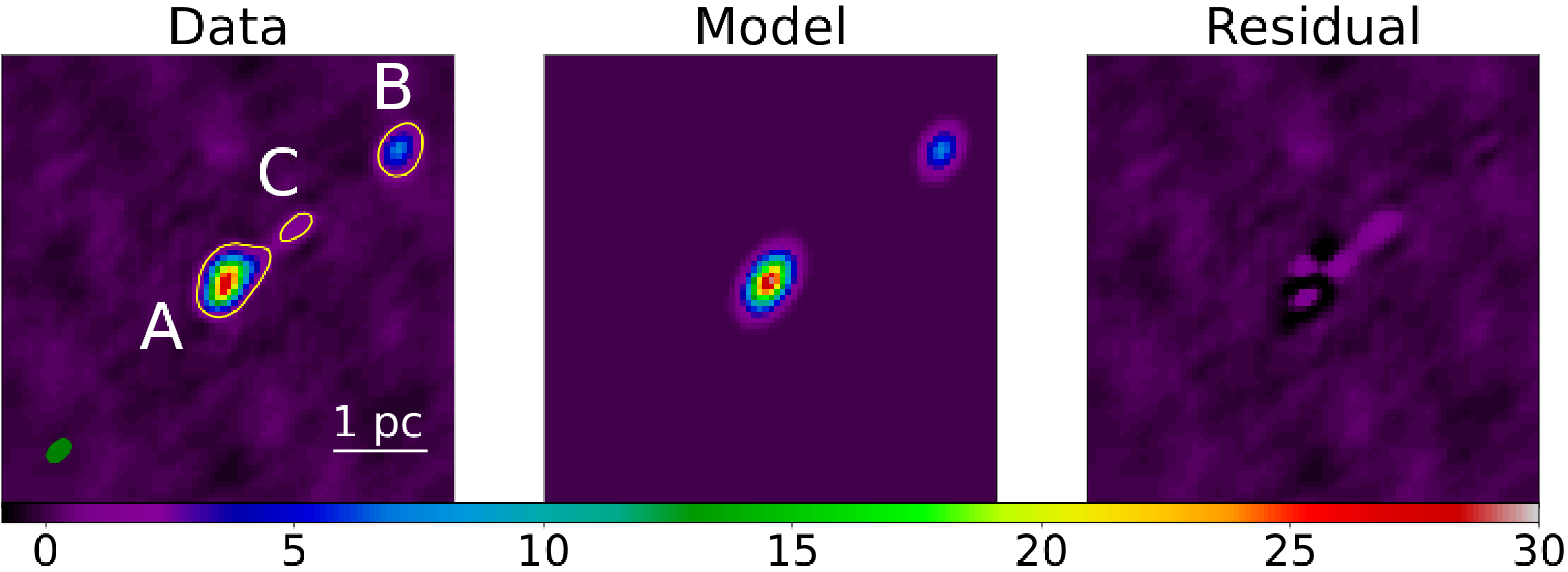}{\textwidth}{}}\vspace{-0.7cm}
    \gridline{\fig{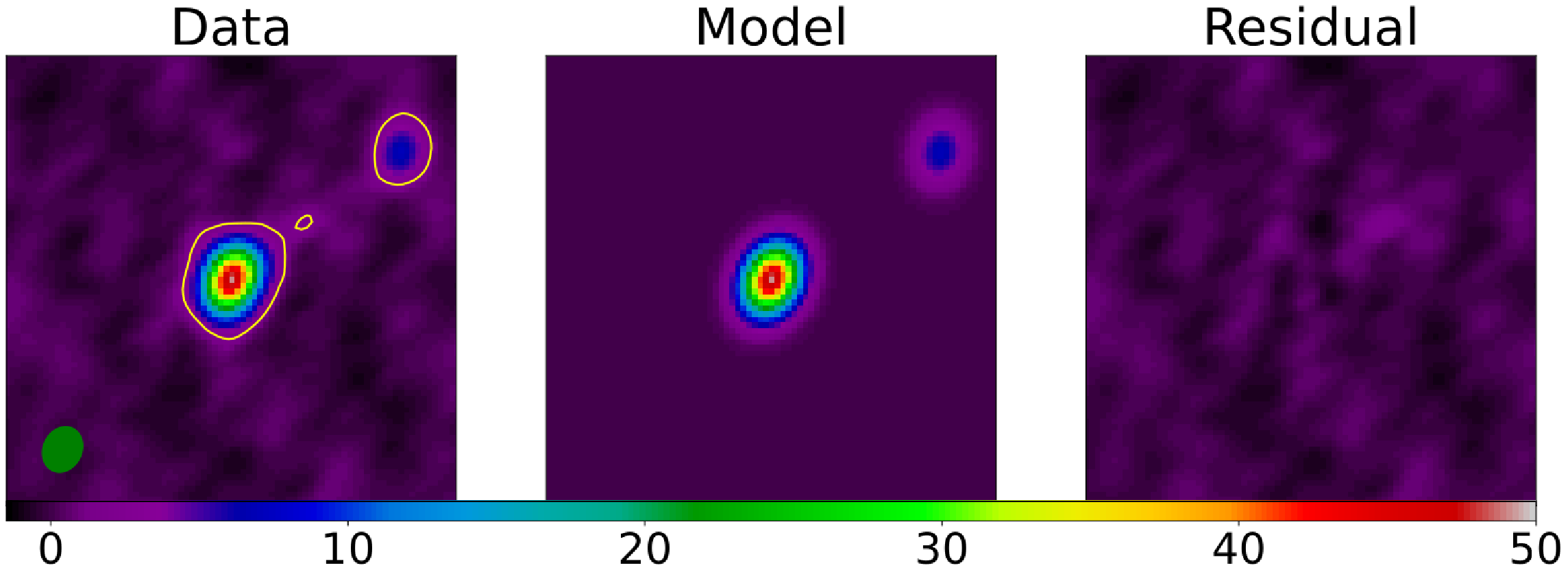}{\textwidth}{}}\vspace{-0.7cm}
    \gridline{\fig{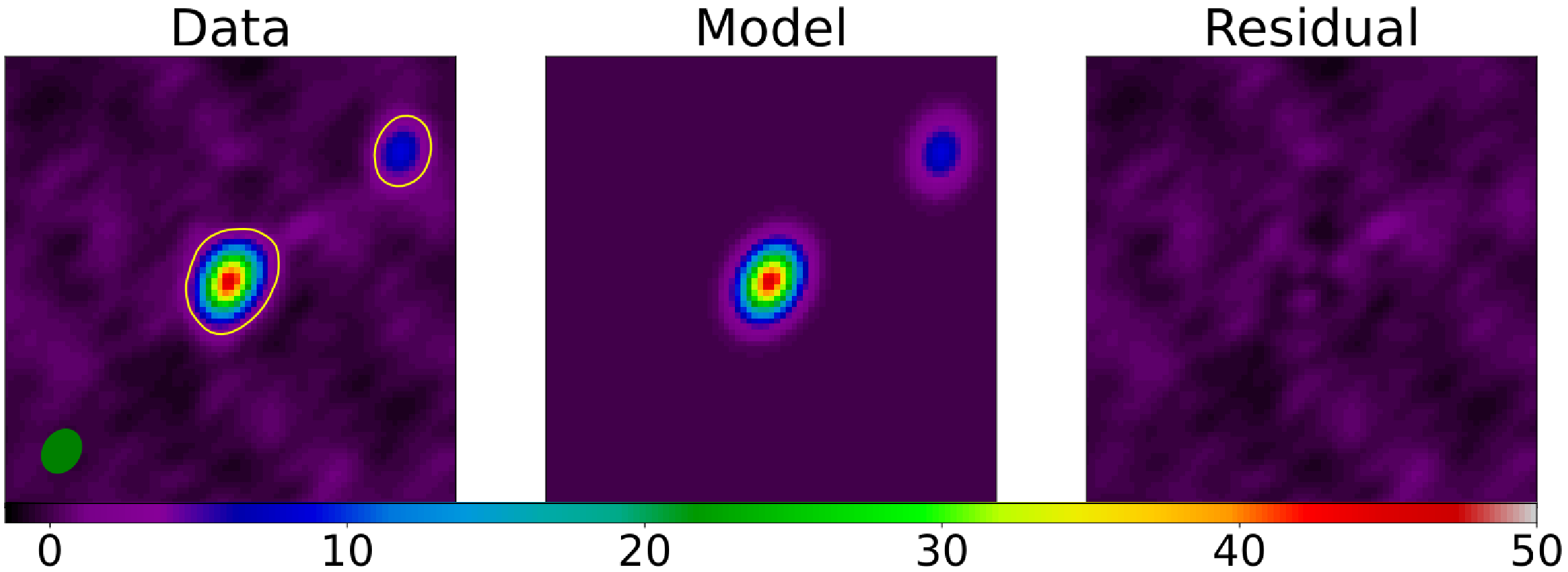}{\textwidth}{}}\vspace{-0.7cm}
    \figcaption{From our \edit1{6~GHz VLBA observations using our }model fitting of radio core components \textit{A} and \textit{B}. Images are $64\times64$~mas, centered on the core component \textit{A}. Color bar is \edit1{in units of mJy beam$^{-1}$.} Green ellipse is the convolved beam. Top Row: Un-tapered observation on 19 May 2020 with included 15~$\times \ \mathrm{RMS}$ contours. Both middle and bottom row images tapered to 48 M$\lambda$ in the \edit1{$uv$} plane to constrain data as close to a circular convolved beam with included 10~$\times \ \mathrm{RMS}$ contours. Middle Row: Observation on 27 Jan 2020. Bottom Row: Observation on 19 May 2020.
    \label{fig:data_model_residual}}
\end{figure*}

\subsubsection{Analysis}

Similar to \citet{2022ApJ...927...18F}, we used AIPS to calculate the RMS noise with the task \textsc{imean}, then used the task \textsc{jmfit} to calculate the peak and integrated flux for components \textit{A}, \textit{B}, and \textit{C}. Table~\ref{tab:radio_measurements} lists our measurements.

To determine the separation between components \textit{A} and \textit{B} and minimize the uncertainty in their position, we needed to create additional images that had a convolved beam with equal major and minor axes, similar to what was done in \cite{2007MNRAS.377..731M}. This is due to detecting complex source structures that are resolved, making it problematic to accurately measure source positions and their separation (see top of Fig.~\ref{fig:data_model_residual}). To account for source extent we used a taper of 48~M$\lambda$ in the $uv$ plane to obtain a nearly circular convolved beam while making sure to use as much of the data as possible and avoiding source confusion. To measure the separation, we constructed Monte Carlo simulations to estimate the posterior distribution of the distance between components \textit{A} and \textit{B}. This was done by performing an initial fit of a two-dimensional, elliptical Gaussian function to each component (Figure~\ref{fig:data_model_residual}). With this fit as a source model, we produced background images with correlated noise by convolving an elliptical Gaussian beam kernel with uncorrelated noise and adjusting the RMS of this noise to match that of the data after convolution. We first compared the separation of \textit{A} and \textit{B} based on our initial fit from the elliptical Gaussian with that of the measured separation between peak brightness of \textit{A} and \textit{B}. For both observations, we obtained a difference of $<$~0.3 pixels ($<$~0.24 mas).
    
While the source positions from the elliptical Gaussian fits are therefore consistent with the peak positions within the uncertainties of the latter, we nonetheless considered pixel-to-pixel position uncertainty in our Monte Carlo simulations by adding an independent random value between -0.5 and 0.5 to the $x$ position and to the $y$ position of the modeled fit. We repeated this $10^5$ times, at each iteration adding correlated noise to the source model and re-fitting each source. We also found an archival image from the VLBA Imaging and Polarimetry Survey \citep[VIPS;][]{2007ApJ...658..203H} in which NGC~3079 was observed on 19~June~2006 at 5~GHz. We repeated our Monte Carlo method using this image and found that the difference between the source separation using the fit model and that using the peak-to-peak separation was also $<0.5$~pix which was consistent with our tapered data. We list our source separation measurements, along with their \edit1{90~\%} confidence intervals (CIs), in Section \ref{subsec: distance of bullet}.        

\subsection{X-ray observations}
We requested observation time using the Swift X-Ray Telescope (XRT) with Target of Opportunity (ToO) (PI: N. Secrest) to be simultaneous with our VLBA observations. The XRT has a PSF with half power diameter of $18\arcsec$ at 1.5~keV with a positional accuracy of $3\arcsec$ and observes at an energy range from $0.2-10$~keV. We requested an integration time of 1.8~ks using Photon Counting (PC) mode and generated the X-ray spectra using the online XRT product generator \citep{2009MNRAS.397.1177E}, setting the source extraction coordinates to the location of the VLBA sources. As described in FRAMEx~II, we were unable to obtain observations for February 2020 and April 2020. In any case, we were not able to extract enough counts for X-ray spectral fitting, likely because the AGN in NGC~3079 is Compton-thick. Fortunately, we found a 24~ks NuSTAR observation taken serendipitously within a month of our 19~May~2020 VLBA observation, providing a quasi-simultaneous constraint on the $3-79$~keV X-ray luminosity of NGC~3079.

\subsubsection{X-ray Analysis}
As in \citet{2022ApJ...927...18F}, we used \textsc{xspec} \texttt{v.12.11.1} \citep{1996ASPC..101...17A} software to perform the spectral analysis. Since the AGN in NGC~3079 is Compton-thick, we used a physically self-consistent fitting model instead of a phenomenological combination of different spectral components. Specifically, we used MYTorus \citep{2009MNRAS.397.1549M} to fit the X-ray spectrum. To robustly estimate the model errors and covariances, the command \texttt{chain} was used to produce Monte Carlo Markov Chains (MCMC) allowing us to obtain \edit1{90~\%} CI for our model's free parameters. Using the MCMCs to estimate the posterior distributions of the constituent model components (e.g., the power-law continuum), we calculated the intrinsic X-ray flux and its uncertainty. Our results are given in Section~\ref{sec: X-ray results}.

\section{Results} \label{sec: Results}

\begin{figure}
    \centering
    {19 June 2006}\vspace{-0.2cm}
    \gridline{\fig{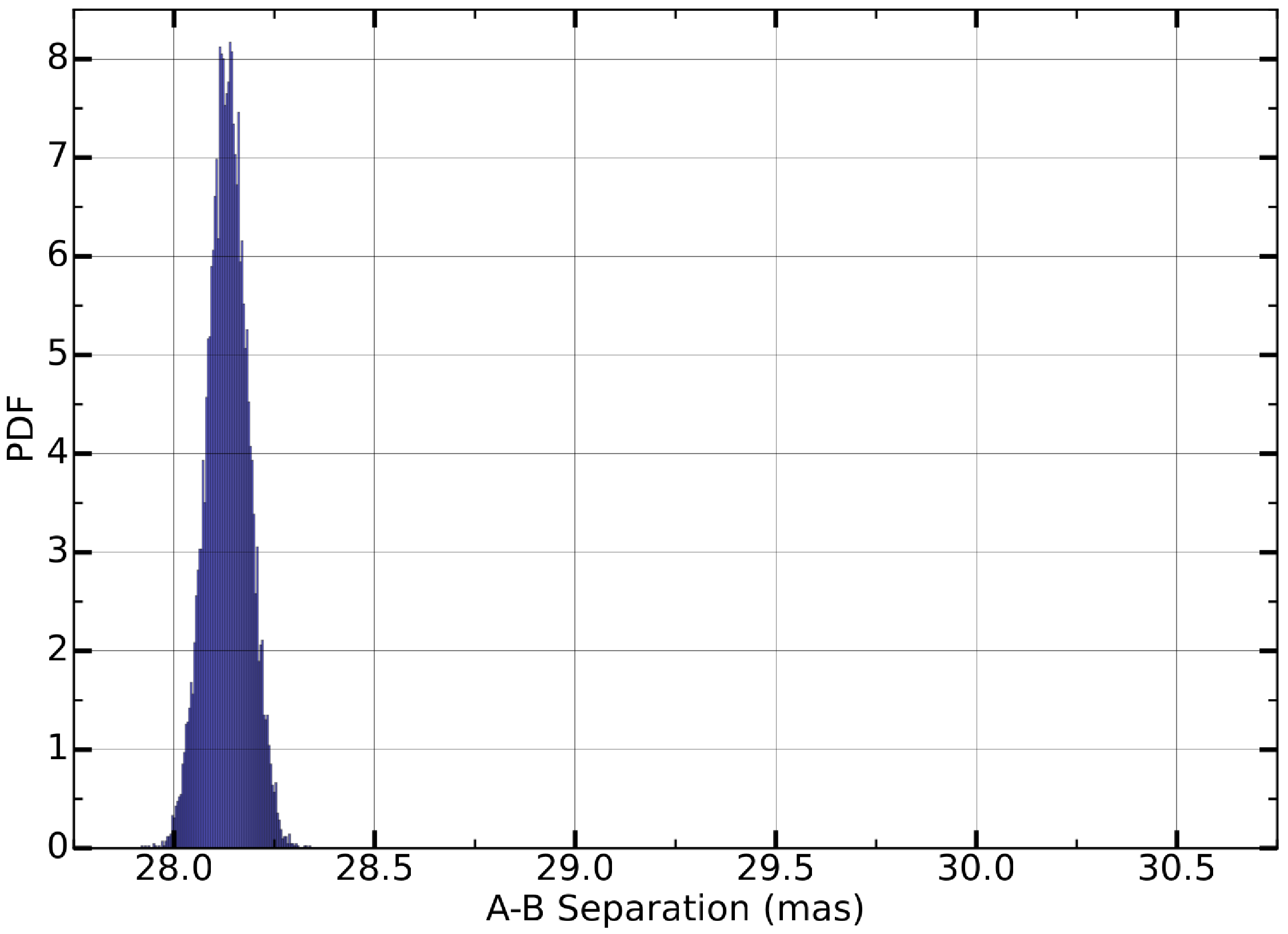}{\columnwidth}{}}\vspace{-0.6cm}
    {27 January 2020}\vspace{-0.25cm}
    \gridline{\fig{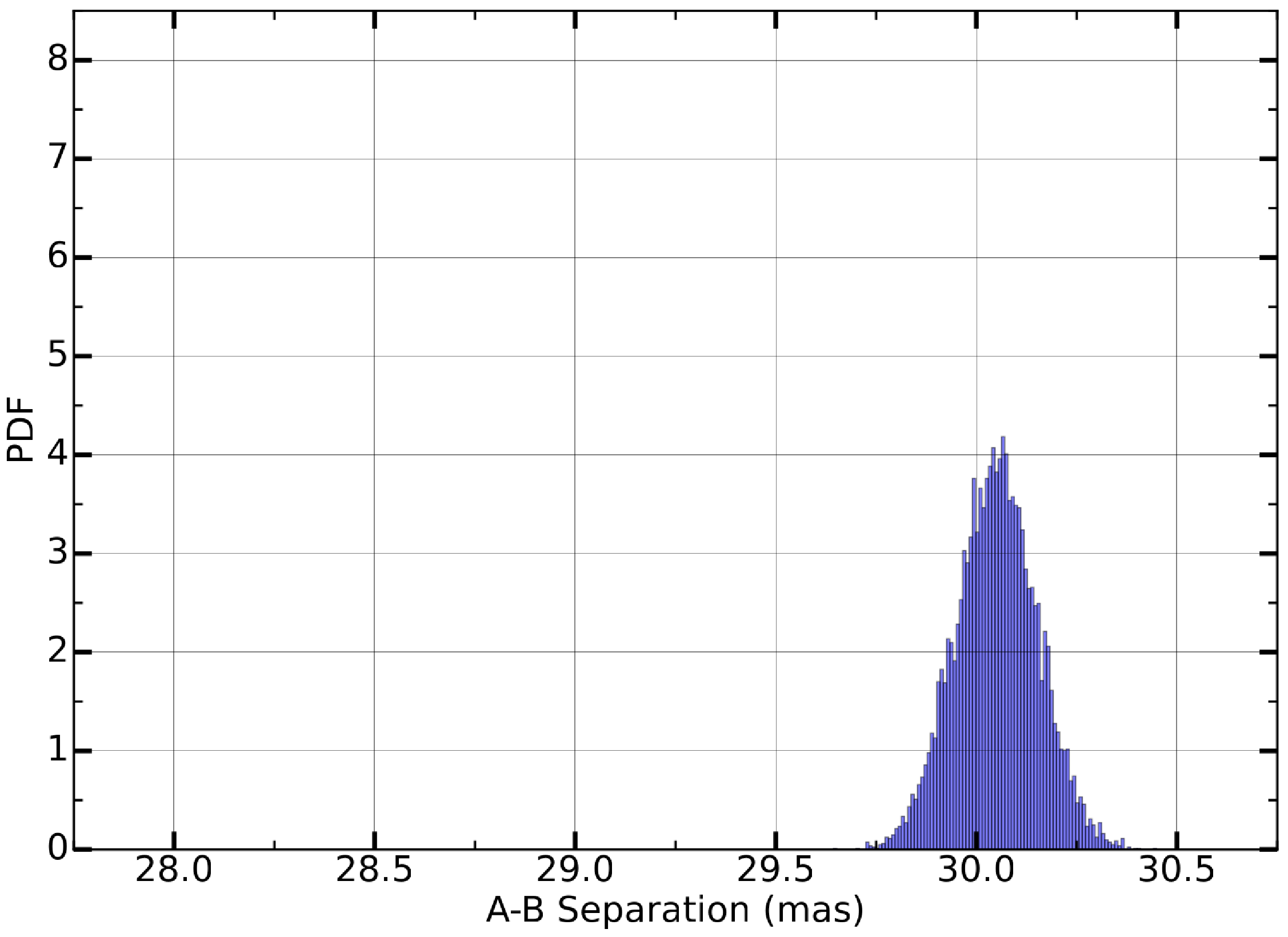}{\columnwidth}{}}\vspace{-0.6cm}
    {19 May 2020}\vspace{-0.25cm}
    \gridline{\fig{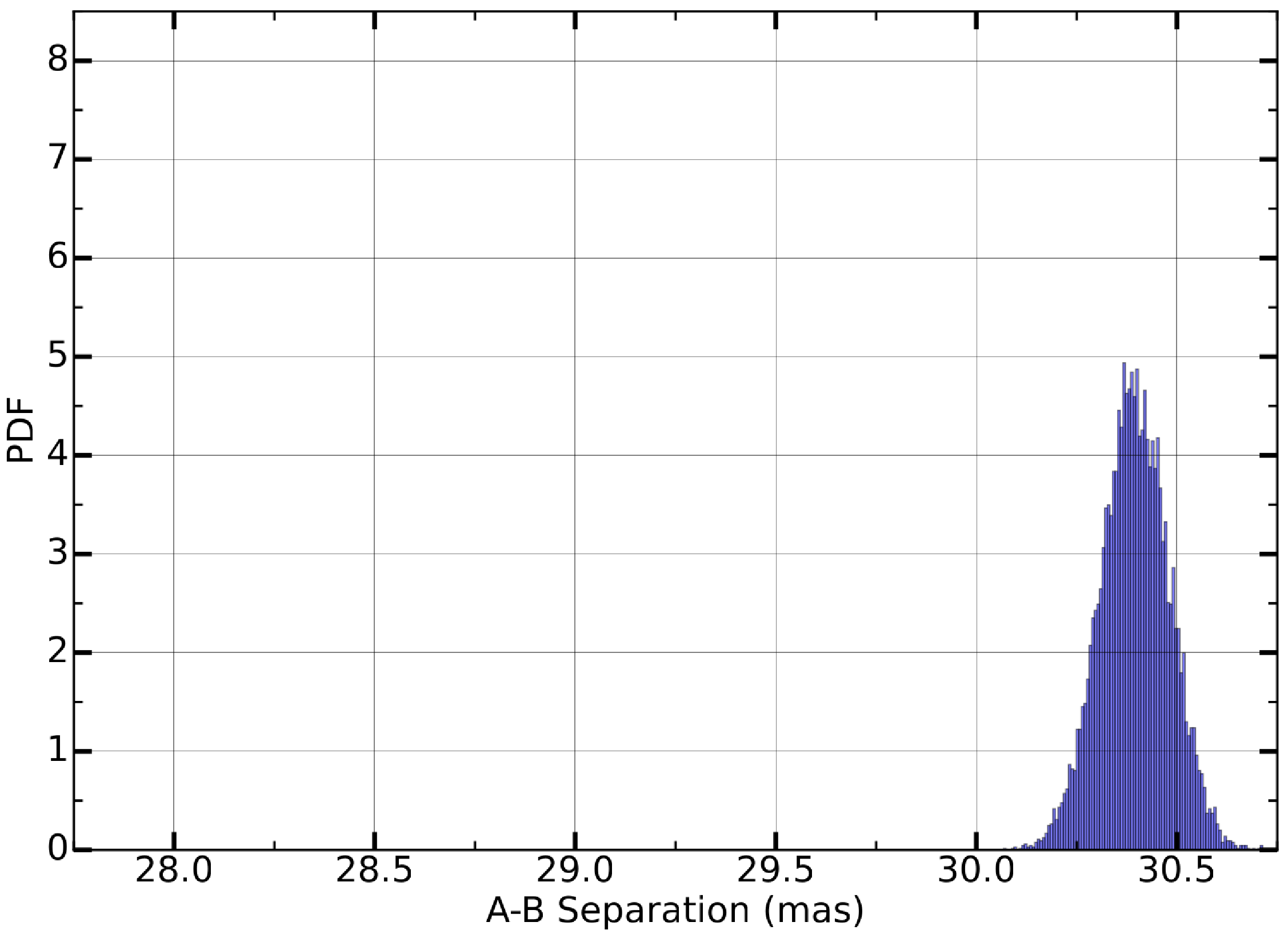}{\columnwidth}{}}\vspace{-0.6cm}
    \figcaption{\label{fig:rposterior}Posterior distribution of projected offset between nuclear components \textit{A} and \textit{B} in NGC~3079. Top: \edit1{For our analysis of the data from \citet{2007ApJ...658..203H}.} Middle: For our observation on 27 Jan 2020. Bottom: For observation on 19 May 2020. 
    }\vfill\null

\end{figure}

\subsection{Separation of Radio Components A and B} \label{subsec: distance of bullet}

\begin{figure}
    \centering
    \gridline{\fig{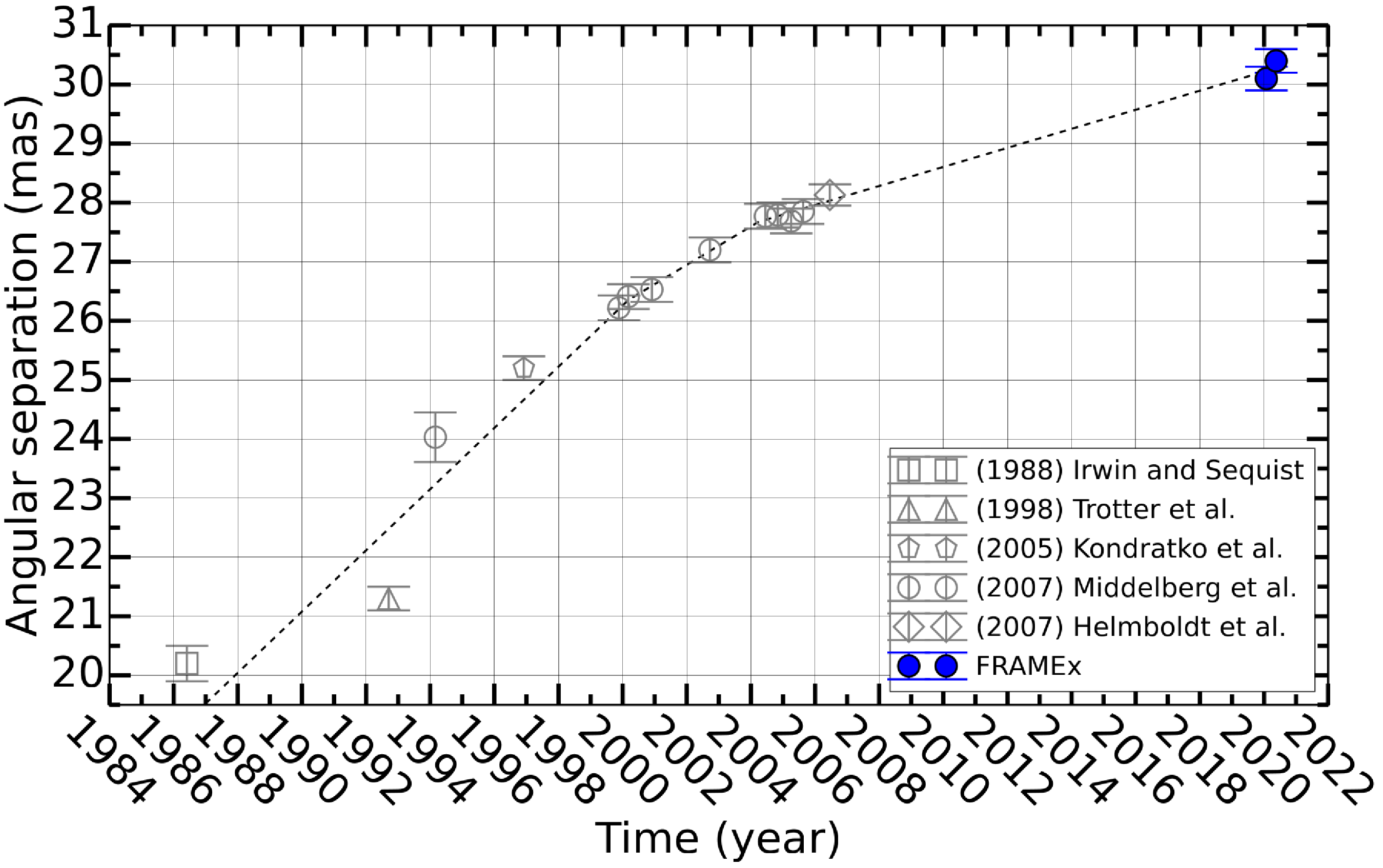}{\columnwidth}{}}\vspace{-0.7cm}
    \gridline{\fig{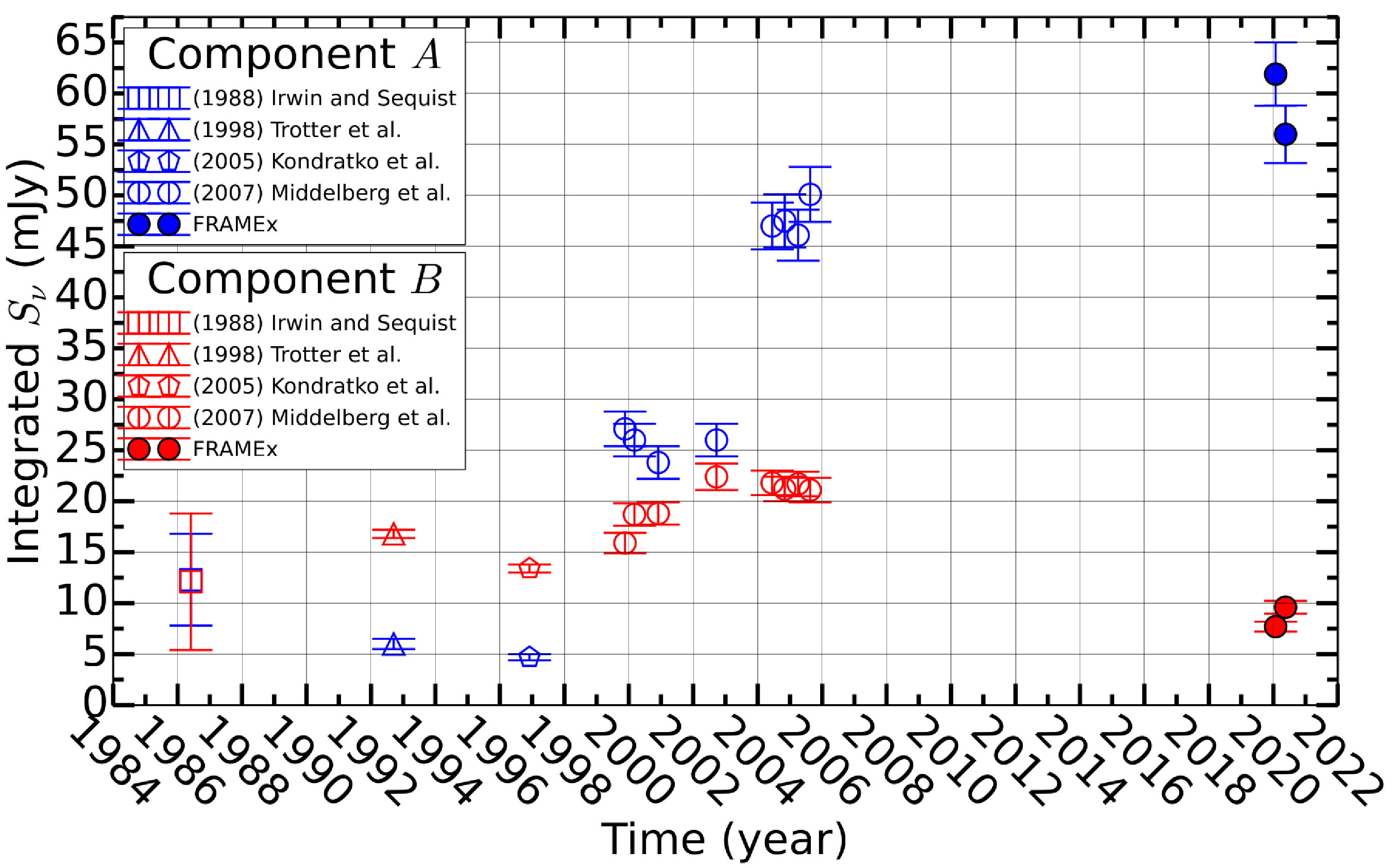}{\columnwidth}{}}\vspace{-0.7cm}
    \figcaption{\label{fig:Seperation}5~GHz archival data points are hollow markers while our 6~GHz data are filled circles. Top: Separation of core components A and B from NGC~3079. Black dashed line indicates a broken linear fit with 3 slopes corresponding to the separate brightening. Bottom: Integrated flux over time.} 
    
\end{figure}

The results for the observation on 19~June~2006 from VIPS gave a separation between nuclear components \textit{A} and \textit{B} of $28.13\pm0.09$~mas, where the uncertainty is \edit1{90~\%} CI and derived from the posterior distribution. In our recent 27~Jan~2020 observation, the separation is $30.1\pm0.2$~mas, while in our 19~May~2020 data it is $30.4\pm0.2$~mas. We show the posteriors from these latter two observations in Figure~\ref{fig:rposterior}. We plot the separation over time, including the 5~GHz separation measurements from \citet{1988ApJ...335..658I}, \citet{1998ApJ...495..740T}, \citet{2005ApJ...618..618K}, and \citet{2007MNRAS.377..731M}, in the top panel of Figure~\ref{fig:Seperation}, along with their corresponding integrated flux values in the bottom panel. It is immediately clear that the slow-down of the $A-B$ separation reported in \citet{2007MNRAS.377..731M}, where they measured \edit1{projected velocities} of ($0.12\pm0.02$)c, ($0.08\pm0.01$)c, and ($0.01\pm0.02$)c corresponding to two periods of slowdowns, has continued to the present day, and that there are step-like jumps in the brightness of component \textit{A} contemporaneous with this slow-down. Given the sampling of the data and its uncertainties, a polynomial fit is not statistically justified, so we instead use a broken linear fit, with a linear component for each jump in the brightness of source \textit{A}. The components of the this linear fit have projected velocities of $(0.13\pm 0.01)$c, $(0.08\pm 0.03)$c, and $(0.040\pm 0.003)$c \edit1{(alternatively, $(0.52\pm 0.02)\mathrm{mas}~\mathrm{yr}^{-1}$, $(0.3\pm 0.1)\mathrm{mas}~\mathrm{yr}^{-1}$, and $(0.16\pm 0.01)\mathrm{mas}~\mathrm{yr}^{-1}$)} corresponding to the period before 2000, 2000--2004, and 2004--present, respectively. \edit1{If we examined only the} angular separation over time, there is a possibility that only one slowdown took place and fitting a two component broken linear fit could be justified. When comparing the $\chi^{2}_{Red}$ between the \edit1{two- and three- component fits}, this resulted in a value of 4.34 and 4.19 respectively. \edit1{Given the previous physical changes seen in component \textit{A} (two jumps in brightness that were co-temporal with the apparent slow downs) along with providing a better $\chi^{2}_{Red}$, this provided enough evidence in favor of the three-component fit and was used to calculate the projected separation rate of the radio components \textit{A} and \textit{B}.} Finally, the brightness temperatures of \textit{A} and \textit{B} are $T_{\text{b}} \sim 10^{9}$~K and $T_{\text{b}} \sim 10^{8}$~K, respectively, indicating \edit1{compact synchrotron emission.}

\subsection{Radio-Loudness}\label{sec: X-ray results}

After fitting the NuSTAR data using the physical MYTorus model and obtaining the Markov chains, the best-fit normalization of the intrinsic power-law is 6$^{+6}_{-3}\times 10^{-3}$ photons~keV$^{-1}$~cm$^{-2}$~s$^{-1}$ at 1 keV, with a photon index of $\Gamma=1.6^{+0.1}_{-0.1}$, and a neutral hydrogen column density of $N_\mathrm{H}=1.8^{+0.2}_{-0.2}\times 10^{24}$~cm$^{-2}$ (all uncertainties correspond to the \edit1{90~\%} CIs). The corresponding 2--10~keV X-ray flux is $3^{+2}_{-1}\times 10^{-11}$~erg~s$^{-1}$~cm$^{-2}$ before absorption, which gives a luminosity of $9^{+2}_{-1}\times 10^{41}$~erg~s$^{-1}$ or log$(L_X$ / erg s$^{-1})\sim42$. 

When it comes to determining the ratio for the radio-loudness parameter, there is a source of ambiguity. The radio-loudness parameter was derived using AGN core radio luminosities at different spatial scales. The C-band observations with the VLA are at a resolution of $\lesssim1\arcsec$, which is unable to resolve any of the core nuclear components, while the VLBA observations are at a resolution of $\sim2$~mas, which resolves out any extended emission and may contain multiple nuclear radio components. Additionally, there is ambiguity as to what constitutes the radio core, where at mas scales the apparent position of the core is frequency-dependent because of synchrotron self-absorption \citep{2019MNRAS.485.1822P} which appears as a `core shift'. Therefore, depending on which spatial scale is used, this can drastically change the result of $R_{X}$. A majority of the sources used in determining the radio-loudness parameter of -4.5 in \citet{2003ApJ...583..145T} were from VLA observations (11 objects) and a small minority were from VLBA observations (4 objects) where all sources are within 60~Mpc. In other words, at these distances, \edit1{the VLA is probing both the parsec scale (compact) and kpc scale (extended) radio emission.} This means that if there are multiple nuclear compact radio sources, the VLA is unable to resolve them out and is therefore observing the sum of all radio emission that can be detected by the VLA's convolved beam. Since a large statistical representation of the sources used in determining the radio-loudness parameter log$(R_X) > -4.5$ is from VLA observations, we therefore summed the luminosities from each nuclear component observed by the VLBA (\textit{A} and \textit{B} from our tapered images) and find log$(R_X)=-4.0$. By the log$(R_X)>-4.5$ radio-loudness threshold established in \cite{2003ApJ...583..145T}, NGC~3079 is therefore radio-loud. Similarly, using a VLA observation, \edit1{\cite{2019MNRAS.485.3185C} found the ratio log$(R_X)=-4.4$, which would also imply that NGC~3079 is radio-loud} given the delineation of -4.5. For the purpose of this analysis, the exact demarcation of what is considered radio-loud vs radio-quiet is not as important as the relative differences between $R_{X}$ (measured in the same way) for the full FRAMEx sample. In other words, those objects with a higher $R_{X}$ are more likely to be jet-powered than those with lower $R_{X}$ value even if $R_{X}$ is biased or ambiguous in some way. 

There is evidence that suggests, based on the presence of water maser emission, that the true position of the central engine lies between components \textit{C} and \textit{B} \citep[][]{1998ApJ...495..740T,2005ApJ...618..618K} suggesting that the radio emission is due to the presence of a jet. Figure~6 in \citet{2005ApJ...618..618K} shows what is described as maser emission tracing a nearly edge-on molecular disk at the position of the AGN. In \citet[][]{2007MNRAS.377..731M}, they found the 5~GHz to 15~GHz spectral index of component \textit{A} changed from steep spectrum to flat spectrum over 3 observations from 1999 to 2000 ($\alpha=1.07$, 0.90, and 0.27 respectively, with $S_{\nu} \propto \nu^{-\alpha}$) while component B remained inverted ($-1.02$, $-0.97$, and $-1.00$ respectively).

It is therefore likely that the central engine lies somewhere between radio knot \textit{A} and \textit{B}, which are currently separated by a projected distance of $\sim2$~pc. If NGC~3079 were at a higher redshift or observed at a lower spatial resolution, \edit1{such as with the VLA}, it would be \edit1{indistinguishable from} that of an AGN core. \edit1{This means the radio emission in radio-loud AGNs may not correspond solely to the central engine (true core) and be mostly due to the effect it has on the surrounding medium due to feedback mechanisms. This would mean the radio emission observed in NGC~3079 does not correspond to the central engine and also does not preclude it from being validly classified as radio-loud.} However, \cite{2019MNRAS.485.3185C} suggests that the log$(R_X) >-4.5$ criterion is too low and a better parameter to follow is a ratio log$(R_{X})=-2.755$ found by \cite{2007A&A...467..519P}. Additional evidence that suggests this may be the case is presented in \cite{2022ApJ...936...76S}. In general, the radio-loudness of an object can be used to estimate if its radio emission is powered by a relativistic jet if the jet itself is difficult to distinguish, due to distance, lack of resolution, or a lack of beaming. As the hard X-ray emission is produced nearest to the SMBH in the compact corona, the VLBA probes a spatial scale more causally connected to the immediate accretion rate inferred from the hard X-rays. Radio emission within the sub-parsec to parsec scales resolved by the VLBA therefore provides a more co-temporal picture of the ratio between the mechanical output of the AGN and its radiative output. Regardless of which $R_X$ criterion is used, the radio-loudness of NGC~3079 compared to the rest of the volume-complete FRAMEx sample suggests that its nuclear radio emission is jet-driven.

\section{Discussion} \label{sec: Discussion}

When we compared our measured separation rate of NGC~3079's nuclear core components \textit{A} and \textit{B} with \cite{2007MNRAS.377..731M}, we find the slow down is not as extensive as previously measured. Their measured separations from 2004 to 2006 are within each other's uncertainty, meaning their measured separation rate was limited by the short spacing between observations. This caused their calculated separation rate to appear slower than the actual separation. Since our new observations took place over 10 years after the previously measured observations, we provide a more accurate depiction of how components \textit{A} and \textit{B} are separating. It is clear based on the changes in luminosity and spectral index of component \textit{A} to a more flat spectrum \citep{2007MNRAS.377..731M}, \textit{A} has impacted a more dense region of the surrounding ISM. Meanwhile, component \textit{B} has retained an inverted spectrum while the luminosity has reduced over time. This indicates only \textit{A} has undergone a recent change in its interaction with the ISM. 

\subsection{Mechanical Feedback} \label{subsec: Mechanical Feedback}
The apparent motion of source \textit{A}, indicated both by the increasing $A-B$ separation and the dramatic brightening of \textit{A} associated with a slowing of the $A-B$ separation, implies AGN-driven kinematics, in which kinetic energy from a jet or wind is deposited into the ISM. There are generally two modes of mechanical feedback: jet-mode and wind-mode, the latter being the uncollimated deposition of kinetic energy from an energetic particle accretion disk wind, but the degree of collimation varies and there may be large, relatively compact masses of material ejected from the AGN accretion disk in discrete events, analogous to the coronal mass ejections seen in stars \citep[e.g.,][]{2022ApJ...927...18F}. 

Given the compact nature of \textit{A}, we quantitatively assess the jet and compact wind-driven mass scenarios here. In the jet scenario, source \textit{A} is a shock front propagating away from the AGN as the jet drills into the ISM. The collimated jet itself, not being highly beamed, is much less luminous, although there is a faint possibly linear feature connecting source \textit{C} to source \textit{A} visible in the residual image of Figure~\ref{fig:data_model_residual} that supports the existence of a jet, as is predicted if radio-loudness is a hallmark of jet activity. The synchrotron emission in source \textit{A} is generated from in~situ production of relativistic electrons by the jet shocking the ISM, unlike the compact mass scenario in which a specific blob of material is physically propagating away from the AGN.

The average power in synchrotron radiation emitted by a single, relativistic electron with an isotropic pitch angle distribution is \edit1{\citep[as reviewed in][]{2016era..book.....C}}:

\begin{equation}\label{eq: power}
    \langle P \rangle = \frac{4}{3}\sigma_{T}\beta^{2}\gamma^{2} c U_{B}
\end{equation}

\noindent where $U_{B}= \frac{B^{2}}{8\pi}$ is the magnetic energy density, $\sigma_{T}$ is the electron Thompson cross section, $\gamma$ is the Lorentz factor, and $\beta \equiv \frac{v}{c}$. For synchrotron emission from an electron, we estimated $\gamma$ using

\begin{equation}\label{eq: gamma}
    \gamma \approx \left( \frac{2 \pi m_{e} c \nu}{e B} \right)^{1/2}
\end{equation}

\noindent where $\nu$ is the observed frequency \citep{2016era..book.....C}. \cite{2019ApJ...883..189S} calculated the minimum magnetic field strength of the host disk of NGC~3079 by fitting models with different volume filling factors to their observational data. They found a magnetic field strength of 35.9~$\mu \mathrm{G}$ for a filling factor of 1, 43.8~$\mu \mathrm{G}$ for a filling factor of 0.5, and 500~$\mu \mathrm{G}$ for a filling factor of $10^{-4}$. Using these magnetic field strength estimates in Equation~\ref{eq: gamma}, in order for an electron to produce synchrotron emission, $\gamma$ must be approximately equal to 7600, 6880, and 2040 respectively. These values appear to be appropriate given relativistic simulations that reproduce synchrotron emission at radio frequencies where the electron Lorentz factors used are between 10 to $10^{5}$ \citep[$\gamma\sim10^4$ used in][]{2019ApJ...885...80N,2021MNRAS.508.5239Y}. After combining Equations \ref{eq: power} and \ref{eq: gamma} with the estimated magnetic field strengths, we found the power from each electron to be $7.9 \times 10^{-17}$~$\mathrm{erg} \ \mathrm{s}^{-1}$, $9.6 \times 10^{-17}$~$\mathrm{erg} \ \mathrm{s}^{-1}$, and $1.1 \times 10^{-15}$~$\mathrm{erg} \ \mathrm{s}^{-1}$ respectively. Taking the ratio of the observed luminosity with the calculated power for an electron gives an estimate to the number of electrons needed in order to produce the luminosity observed. With the estimated number of electrons for different magnetic field strengths, we are able to estimate a range of mass of the shocked electrons in component \textit{A} to be between $M_e = 4 \times 10^{-8}$~M\textsubscript{\(\odot\)} and $M_e = 5 \times 10^{-7}$~M\textsubscript{\(\odot\)}. 

Using the estimated total electron mass, we calculated the mechanical energy deposited into the interstellar medium by the jet acting on component \textit{A}. As the shocks act on in~situ electrons, the initial average net velocity of the electrons can be assumed to be $\sim0$, so the relativistic motions of the electrons along magnetic field lines is powered by the mechanical energy deposited on the ISM by the jet. Then, the total kinetic energy, expressed as mechanical power is:

\begin{equation} \label{eq: kinetic}
P_K = 2M_e c^2 (\gamma - 1)
\end{equation}

\noindent where $M_e$ is the total mass of shocked electrons as calculated above and $\gamma$ is their Lorentz factor. We treat the ISM as having an equal number of protons and electrons in bulk, where the protons absorb as much kinetic energy from the jet on average, giving the factor of 2. Given the range of mass from the magnetic field estimates, the mechanical power is between $3\times 10^{50}$~erg and $1 \times 10^{52}$~erg. This is within an order of magnitude of the total average kinetic energy from supernovae ($\sim10^{51}$~erg). Based on the magnetic field estimates, the lifetime of synchrotron radiation is larger than $>10^4$~yr \citep[][]{1992ARA&A..30..575C}, which implies that the energy calculated here is the cumulative amount of kinetic feedback from the AGN given the apparent motion of the shock and its proximity to the AGN. Because we are probing the AGN with the VLBA there is undoubtedly some larger-scale emission that is resolved out. This can be seen when we compared the luminosity of previous \edit1{VLA A-configuration observations at 5~GHz} from the literature \citep[provided in Table~4 of][]{2021ApJ...906...88F} with the total luminosity from all nuclear components added together from our VLBA observations, the ratio between the two is $1.25$. 

As mentioned earlier, there is also the possibility that component \textit{A} is hot plasma ejected from the AGN that has been continuously propagating away from the central engine and collided with a denser region of the ISM during the epoch~$\sim2000-2006$ slow-down/brightening period (Figure~\ref{fig:Seperation}). In this picture, the faint possACKNOWLEDGMENTSibly linear feature could be the trail of the ejected hot plasma from previous shocks, and the kinetic energy difference $\Delta$KE captures how much energy is being dumped into the surrounding ISM. We calculated the relativistic $\Delta$KE by using the projected separation velocity from before and after the collision, with the mass estimates of component \textit{A} assuming that it is composed purely of electrons. This resulted in a range of $\Delta$KE between $2 \times 10^{44}$~erg and $1 \times 10^{45}$~erg. In reality, component \textit{A} is most likely composed of both electrons and protons, but only the electrons are responsible for synchrotron emission. These numbers are therefore a lower limit on the total amount of kinetic feedback component \textit{A} is having on the surrounding ISM. For the case where there is an equal number of protons and electrons that make up component \textit{A}, the $\Delta$KE range is from $3 \times 10^{47}$~erg to $1 \times 10^{48}$~erg. Since determining the exact composition of component \textit{A} is difficult, there is also the possibility component \textit{A} contains more protons than electrons. If this is the case, the range $2 \times 10^{44}$~erg to $1 \times 10^{48}$~erg is a better lower limit to the total amount of kinetic feedback \textit{A} could be having on the surrounding medium.

\subsection{Comparing Feedback Mechanisms}
In order to better understand and contextualize the effects of AGN-driven, jet-mode feedback on host galaxies, we can leverage the fact that NGC~3079 is part of a volume-complete sample to roughly assess the overall importance of jet-mode, wind-mode, and radiative-mode feedback in the interactions between AGNs and their host galaxies, at least in the local universe.  
Given the considerations previously discussed, we favor the jet interpretation for the radio emission in NGC~3079. This being the case, NGC~3079 joins NGC~1052 as the jet-driven, radio-loud members of the FRAMEx sample. For NGC~1052, we scaled the KE of NGC~3079 (KE from weakest magnetic field strength estimate 35.9~$\mu \mathrm{G}$) by the ratio of their 6~GHz radio luminosities, a factor of 17. Applying this to the mechanical power in component \textit{A}, we find the value for NGC~1052 to be $\sim 10^{53}$~erg, which is within an order of magnitude of the value $(1.3\pm0.9)\times10^{53}$~erg recently found by \citet{2022A&A...664A.135C} using MUSE data. Then, the mean mechanical energy from jet-mode feedback across the FRAMEx sample is $\sim 10^{51}$~erg. This is equivalent to the total average KE from \edit1{a} supernova.

The energetics of wind-mode and radiative feedback can be estimated from the bolometric luminosities of the full FRAMEx sample, which can in turn can be estimated from the hard X-ray luminosities. We used Equation~1 in \cite{2023MNRAS.518.2938T} with the 14--195~keV luminosities listed in \citet{2021ApJ...906...88F} to estimate the bolometric luminosities of the FRAMEx sample. With the estimates of the jet-mode mechanical feedback from NGC~3079 and NGC~1052 in hand, we can exploit the volume completeness of the FRAMEx sample to assess the relative importance of jet-mode feedback in the local universe. As we have seen, NGC~1052 dominates the jet-mode energy budget, with a total feedback of $\sim10^{53}$~erg. While this amount of energy would take $\sim500$~yr to reach given the bolometric luminosity of NGC~1052, if we compare the summed jet-mode feedback energy from NGC~1052 and NGC~3079, the only two radio-loud objects in this volume, with the summed bolometric luminosities of all the AGNs in the volume ($\sim10^{45}$~erg~s$^{-1}$), we find that it only requires $\sim3$~yr for the total, non-jet bolometric output of the AGNs to reach the same amount of energy as their total, jet-mode output. We have seen that, at least for NGC~3079, the energy from the jet has taken $\sim70$~yr to build in the ISM, given the much longer synchrotron cooling times  \citep[$>10^4$~yr;][]{1992ARA&A..30..575C}. As the radio source in NGC~1052 is obviously much older than $\sim3$~yr, and likely at least as old as the source in NGC~3079, this implies that, on average, the jet-mode energy feedback from AGNs in the local universe is of order a few percent of their total bolometric output, indicating that jet-mode feedback is relatively unimportant. A potential caveat of this statement is that there could be older radio lobes and shock relics on larger scales than our VLBA observations are sensitive to. However, given the relatively compact radio morphologies seen in archival VLA data \citep[Figure~3 in][]{2021ApJ...906...88F}, this appears not to be the case. Defining a volume-complete sample has allowed us to integrate over a statistically representative sample of AGNs, allowing for detailed single-object studies such as our study of NGC~3079 to directly inform our understanding of the wider role of AGNs in their environments and host galaxies.

We also examined the amount of wind power that is generated to compare with our results from jet-mode feedback in order to probe the total mechanical feedback of NGC~3079. In order to estimate this, we started from first principles of accretion disks from \citet{1973A&A....24..337S} and created a surface temperature profile using a $M_{BH} = 10^{6.38} M_{\odot}$ \citep[provided in Table 1 of][]{2021ApJ...906...88F}. Using this temperature profile, we calculated a range of specific luminosities ${L_{\nu}}$ for a range of frequencies. ${L_{\nu}}$ was then normalized using the intrinsic 2500\,\AA\ luminosity estimated from the 2\,keV luminosity using the $\alpha_{\mathrm{OX}}$ relationship from \citet{2010A&A...512A..34L}. The normalized ${L_{\nu}}$ was then used to calculate the amount of kinetic energy imparted on outer disk protons and electrons via Compton scattering, accounting for relativistic effects using conservation of energy. Examining the difference in amount of power lost due to a final scattering off of either a proton or electron, yields the total amount of power of the wind ($P_{\mathrm{wind}}\sim4\times 10^{38}$~erg~s$^{-1}$). Using the initial \edit1{projected separation velocity} of components \textit{A} and \textit{B}, we estimated the amount of time elapsed since the separation began with that of our most recent observation ($\sim 70$~yr). Therefore the amount of energy released from wind mode feedback since the separation began is $\sim 9 \times 10^{47}$~erg. This is inline with the wind driven scenario discussed towards the end of \ref{subsec: Mechanical Feedback}. 

In order to determine the possible mechanisms at work in NGC~3079, we then examined the Eddington ratio ($\lambda_{\mathrm{Edd}} = L_\mathrm{bol}/L_\mathrm{Edd}$). First, in order to calculate the bolometric luminosity of the AGN in NGC~3079, we used the bolometric estimate equation from \cite{2023MNRAS.518.2938T} and the 14-195~keV luminosity from FRAMEx I. Then we calculated the Eddington luminosity for a $\log(M_{BH}/M_{\odot}) = 6.38$ in order to find the Eddington ratio and found $\lambda_{\mathrm{Edd}} \sim 0.02$. As a check, we used the bolometric luminosity $\log(L_{\mathrm{bol}}/L_{\odot}) = 10.03 \pm 0.25$ calculated from MIR \citep{2016MNRAS.458.4297G}. For a $\log(M_{BH}/M_{\odot}) = 6.38$, we found $\lambda_{\mathrm{Edd}} = 0.136$. This means even on the upper end, the $\lambda_{\mathrm{Edd}} < 0.2$ for the AGN in NGC~3079. \citet{2019A&A...630A..94G} suggests an AGN with $\lambda_{\mathrm{Edd}}$ at this regime ($10^{-3} \lesssim \lambda_{\mathrm{Edd}} \lesssim 10^{-1}$) is dominated by an optically thick and geometrically thin accretion disk where \edit1{radiation-driven} disk wind is unable to form and feedback is mainly radiative and jet emission is suppressed compared to lower Eddington ratios. Those AGNs with $10^{-6} \lesssim \lambda_{\mathrm{Edd}} \lesssim 10^{-3}$ are expected to be magnetically dominated with high-velocity collimated radio jets with a low velocity outer wind, while AGNs with $\lambda_{\mathrm{Edd}} \gtrsim 0.25$ are said to be wind-dominated. \cite{2022MNRAS.513.1141Z} also comes to a similar conclusion through two-dimensional simulations, where they studied \edit1{large-scale dynamics} of accretion disk winds driven by line force. They found that for black hole masses $<10^7$~M\textsubscript{\(\odot\)} and a $\lambda_\mathrm{Edd} = 0.3$, \edit1{the strength of the winds kinetic energy is substantially weaker at larger radii and unable to provide a sufficient amount of feedback to affect the host galaxy.} \cite{2018FrASS...5....6M} suggests outflow velocities are related more to the Eddington ratio than black hole mass or luminosity. Where at low accretion rates ($< 0.2$), they propose there is a partly failed wind with a geometrically thin accretion disk. This appears to be consistent with what is observed in NGC~3079 given a $\lambda_\mathrm{Edd} \sim0.02$. Due to the nature of the \edit1{volume-complete sample}, this would indicate all AGNs in the local universe with similar Eddington ratios may also be dominated by other forms of feedback (i.e. radiative feedback). 

There is another object in the FRAMEx I volume-complete sample that has similar characteristics to NGC~3079. The AGN in NGC~1068, at a distance similar to NGC~3079, is also a \edit1{Compton-thick source}, also contains multiple nuclear radio components, and has a similar bolometric luminosity \citep{2021ApJ...906...88F}. \edit1{In a recent publication, \cite{2023ApJ...953...87F} }explores apparent motion in some of the radio knots in NGC~1068, but unlike NGC~3079 they conclude that the radio knots are most likely pseudo-motions due \edit1{to} changes in the densest regions in a much larger, extended radio structure that is otherwise resolved out by the VLBA. This behavior is more likely to be due to AGN winds or radiative feedback, instead of jet-mode feedback as we infer for NGC~3079. The reasons why we do not conclude that a similar process is occurring in NGC~3079 are as follows. First, NGC~3079 has been observed at mas scales multiple times over a span of $\sim40$ years and only radial, linear motion away from the maser-inferred position of the AGN has been seen. This is not the expected behavior for pseudo-motion, in which regions dense enough to be detected by VLBI observations randomly condense out of the extended radio emission, leading to non-radial or tangential motions. Second, we know a priori that NGC~3079 is much more radio-compact on larger scales than NGC~1068. In FRAMEx~I, we found that, while the ratio of VLA (A-configuration) to VLBA C-band integrated fluxes is $\sim3$ (and around 1.25 using the data from this work), the same ratio for NGC~1068 is $\sim400$. In other words, NGC~1068 is over two orders of magnitude more radio-extended than NGC~3079, given that they are at nearly the same distance $(\sim16$~Mpc), consistent with radio emission in NGC~1068 being primarily driven by an uncollimated wind. Additionally, while we find a faint, potentially linear feature along historical trajectory of source~A, no such feature is present in NGC~1068. Finally, only NGC~1068 has a source detected at C-band with the VLBA coincident with its assumed AGN core position, while NGC~3079 is undetected at the AGN position inferred from maser emission, implying different core radio production mechanisms. If NGC~3079 is indeed producing a jet as the evidence suggests, then the apparent radio silence of the core may be due to synchrotron self-absorption\edit1{, free-free absorption, or a combination of both.} Nonetheless, NGC~1068 and NGC~3079 also have similar bolometric luminosities (estimated using their hard X-ray luminosities) and SMBH mass, which means they have similar Eddington ratios. Given the differences discussed above regarding the possible physical mechanisms at play with NGC~1068 and NGC~3079, this is somewhat surprising. One explanation is that the SMBH mass of NGC~3079 is actually larger than we have assumed. Indeed, there is considerable variance in literature estimates of NGC~3079's SMBH mass, ranging between $10^6$ and $10^8$~$M_\Sol$ \citep[$10^{6}$, $10^{8}$, $10^{6}$, $10^{7.2}$, $10^{8.2}$;][]{2005ApJ...618..618K,2017MNRAS.467..540L,2021MNRAS.502.3329G,2022MNRAS.510.5102O,2022ApJS..260...30T}. Given this range of masses, the Eddington ratio could be two orders of magnitude less than that of NGC~1068. If this is the case, it could help explain the differences in physical mechanisms inferred for the two objects and add to the evidence in favor of a jet-mode feedback scenario for the AGN in NGC~3079.

\subsection{Possible Scenarios for the Nuclear Component \textit{B}}
The balance of the evidence is that source~\textit{A} is propagating through the ISM, brightening or dimming as the density of the intervening material changes, driven by a likely jet seen as a faint linear structure connecting \textit{A} to the position of the AGN somewhere near \textit{C}. In this scenario, component \textit{B} does not appear to be moving, a conclusion supported by the relative stability of its luminosity (see bottom of Fig.~\ref{fig:Seperation}). We note, however, that \textit{B} has an inverted spectrum \citep{2007MNRAS.377..731M}, which argues against it being relic emission from an earlier accretion event or supernova. 
One possible explanation is that \textit{B} is the termination point of a frustrated jet. A second, less likely scenario is based on the recent work of \citet{2022A&A...658A.119B}, who examined NGC~1052 and found that the emission gap between the base of the jet had a highly inverted radio spectral index and indicated this was due to an obscuring molecular torus causing free-free absorption. In this case, component \textit{B} may be at or near the base of the jet and is being obscured by the torus. Source \textit{A} would therefore be material from an older ejection from the central engine. We view this scenario as less likely because of the position of water maser emission between \textit{A} and \textit{B}, which suggests that the AGN core lies between these two sources. This was shown in \citet{2005ApJ...618..618K} where they found blue-shifted and red-shifted maser emission that traces a nearly edge on molecular disk about a pc in radius and aligned with the kpc scale molecular disk.

\section{Conclusions} \label{sec: Conclusion}

In this work, we have presented an analysis of the radio knot kinematics in NGC~3079, one of two radio-loud members of the volume-complete FRAMEx sample. Using new VLBA observations, we found the current projected rate of separation between the two main knots \textit{A} and \textit{B} to be $(0.040\pm0.003) c$. This is consistent with previous work, but our observations significantly increase the temporal baseline of VLBI monitoring of NGC~3079. The brightening of component \textit{A} co-temporal with an apparent reduction in its velocity suggests that \textit{A} marks an interaction of the AGN with a more dense ISM. We have considered two scenarios for the nature of the radio knot kinematics: shocking of the ISM by an otherwise radio-faint jet, and bulk motion of radio-emitting plasma ejected from the AGN. Our primary conclusions are as follows:

\begin{enumerate}
    \item If the radio knots are jet-powered, then the cumulative amount of kinetic feedback on the surrounding medium is between $3\times 10^{50}$~erg to $1\times 10^{52}$~erg. This scenario is supported by the presence of a faint linear feature seen in our VLBA observations connecting component \textit{A} to the inferred position of the AGN, and is also predicted by the physical picture in which radio-loudness is a consequence of relativistic jets. If the radio knots are ejected plasma, then the cumulative kinetic feedback is between $2 \times 10^{44}$~erg to $1\times 10^{48}$~erg, depending on the composition of component \textit{A}, vastly lower than in the jet scenario.

    \item Since the projected separation velocity was initially $\sim0.13$c, the appearance of a possible linear feature in the residual image from the model subtracted from the data in Fig.~\ref{fig:data_model_residual}, a log$(R_X)=-4.0$ suggesting that NGC~3079 is radio-loud, the scaled total KE from the jet-mode feedback scenario of the AGN in NGC~3079 to the AGN in NGC~1052 (a well known radio-loud jetted AGN) is consistent with the KE found in the literature, and a $\lambda_\mathrm{Edd} \sim0.02$ suggesting a failed or weak wind that is insufficient in producing the amount of mechanical feedback needed to affect the host galaxy, therefore the evidence points towards the jet-mode scenario.  
    
    \item As NGC~3079 is the only member of the volume-complete FRAMEx sample of local AGNs other than the radio-loud NGC~1052 to exhibit the likely jet-powered radio knot kinematics shown here, our results indicate that the amount of mechanical energy produced by jet-mode feedback and wind-mode feedback with Eddington ratio $\sim0.02$, appear to not be as significant as other modes of feedback (i.e. radiative feedback) for AGNs in the local universe.
\end{enumerate}

Further understanding the effects of mechanical feedback from AGNs can inform the overall importance of feedback and its role on galaxy formation and bulges. This can also inform what role mechanical feedback has on ISM/IGM heating in AGNs. For wind-mode feedback, \citet[][]{2015ARA&A..53..115K} suggests small scale momentum-driven outflows (thermal energy is lost to cooling and only ram pressure is conserved) that interact with a small central part of the bulge sets the critical $M-\sigma$ black hole mass. Where energy-driven outflows (wind shock energy is conserved and shocked wind is expanding adiabatically) are global and expand out to greater scales to produce \edit1{large-scale molecular} outflows that can sweep up and clear the galaxy of gas. As for jet-mode feedback, \citet[][]{2012ARA&A..50..455F} reviewed different types of AGN feedback, where jet-mode feedback can create bubbles of relativistic plasma on either side of the nucleus and affect the surrounding interstellar medium. As part of a volume-complete sample, NGC~3079, along with NGC~1052, therefore provides a direct view into mechanical feedback from AGNs in the local universe.

\vfill\null
\acknowledgments
Part of this work is based on archival data, software and online services provided by the Space Science Data Center $-$ ASI.
The National Radio Astronomy Observatory is a facility of the National Science Foundation operated under cooperative agreement by Associated Universities, Inc. The authors acknowledge use of the Very Long Baseline Array under the U.S.\ Naval Observatory’s time allocation. This work supports USNO’s ongoing research into the celestial reference frame and geodesy.
\vfill\null

\facilities{VLBA, SWIFT, NuSTAR}

\software{
Astropy \citep{2013A&A...558A..33A,2018AJ....156..123A}, 
\textsc{AIPS} \citep{2003ASSL..285..109G},
\textsc{XSPEC} \citep{1996ASPC..101...17A}}

\bibliographystyle{aasjournal}
\bibliography{Fernandez_et_al_2023}{}

\end{document}